\documentclass[a4paper]{article}
\usepackage{spconf}
\usepackage{amsmath}
\usepackage{amssymb}
\usepackage{graphicx}
\usepackage{subfigure}
\usepackage[square,comma,numbers]{natbib}
\usepackage{url}

\newcommand{\eqn}[1]{(#1)}

\newcommand{\fig}[1]{Fig.~#1}

\newcommand{\sectn}[1]{Sec.~#1}

\newcommand{\eg}{\mbox{\it e.g.}}

\newcommand{\fov}{FOV}

\newcommand{\wterm}{\mbox{$w$-term}}

\newcommand{\spcend}{\ensuremath{\:}}
\newcommand{\img}{\ensuremath{{\rm i}}}

\newcommand{\disk}{\ensuremath{{D^2}}}
\newcommand{\vect}[1]{\ensuremath{\mbox{\boldmath ${#1}$}}}

\newcommand{\dx}{\ensuremath{\mathrm{\,d}}}

\newcommand{\sa}{\ensuremath{\omega}}

\newcommand{\el}{\ensuremath{\ell}}

\newcommand{\sind}{\ensuremath{{\rm s}}}

\renewcommand{\exp}[1]{\ensuremath{{\rm e}^{#1}}}

\newcommand{\sao}{\ensuremath{\sa_0}}

\newcommand{\vis}{\ensuremath{y}}

\newcommand{\beam}{\ensuremath{A}}

\newcommand{\im}{\ensuremath{x}}

\newcommand{\nim}{\ensuremath{n}}
\newcommand{\visvect}{\ensuremath{\vect{\vis}}}
\newcommand{\imvect}{\ensuremath{\vect{\im}}}

\newcommand{\impvect}{\ensuremath{\vect{\imp}}}
\newcommand{\imsvect}{\ensuremath{\vect{\ims}}}

\newcommand{\nimvect}{\ensuremath{\vect{\nim}}}
\newcommand{\bu}{\ensuremath{u}}
\newcommand{\bv}{\ensuremath{v}}
\newcommand{\bw}{\ensuremath{w}}

\newcommand{\buvect}{\ensuremath{\vect{\bu}}}
\newcommand{\buvectfull}{\ensuremath{\vect{b}}}

\newcommand{\lx}{\ensuremath{l}}
\newcommand{\mx}{\ensuremath{m}}
\newcommand{\nx}{\ensuremath{n}}
\newcommand{\lxvect}{\ensuremath{\vect{\lx}}}

\newcommand{\chirp}{\ensuremath{C}}
\newcommand{\chirpfull}{\ensuremath{\chirp(\| \lxvect \|_2)}}

\newcommand{\nmeas}{\ensuremath{M}}
\newcommand{\ndim}{\ensuremath{N}}

\newcommand{\ispar}{\ensuremath{i}}

\newcommand{\sparatom}{\ensuremath{\vect{\psi}_\ispar}}

\newcommand{\sparmat}{\ensuremath{\Psi}}
\newcommand{\sensmat}{\ensuremath{\Phi}}

\newcommand{\sensmatp}{\ensuremath{\Phi_{\pind}}}
\newcommand{\sensmats}{\ensuremath{\Phi_{\sind}}}

\newcommand{\opmask}{\ensuremath{\mathbfss{M}}}
\newcommand{\opbeam}{\ensuremath{\mathbfss{A}}}
\newcommand{\opfourier}{\ensuremath{\mathbfss{F}}}

\newcommand{\opchirp}{\ensuremath{\mathbfss{C}}}

\newcommand{\opproj}{\ensuremath{\mathbfss{P}}}

\renewcommand{\fov}{FoV}
\renewcommand{\sa}{\ensuremath{\vect{\hat{s}}}}
\renewcommand{\opmask}{\ensuremath{\mathbf{M}}}
\renewcommand{\opbeam}{\ensuremath{\mathbf{A}}}
\renewcommand{\opfourier}{\ensuremath{\mathbf{F}}}

\renewcommand{\opchirp}{\ensuremath{\mathbf{C}}}

\renewcommand{\opproj}{\ensuremath{\mathbf{P}}}
\renewcommand{\sensmatp}{\ensuremath{\Phi}}
\renewcommand{\impvect}{\ensuremath{\vect{\im}}}
\renewcommand{\imsvect}{\ensuremath{\vect{\im}_{\sind}}}
\renewcommand{\wterm}{\mbox{$w$-modulation}}
\newcommand{\Cs}{CS}
\newcommand{\cs}{CS}
\newcommand{\csfirst}{compressed sensing (CS)}
\urldef\myhomepage\url{www.jasonmcewen.org}

\title{COMPRESSED SENSING FOR RADIO INTERFEROMETRIC IMAGING:\\REVIEW AND FUTURE DIRECTION}
\name{Jason D. McEwen${}^1$\thanks{URL: \myhomepage} \& Yves Wiaux${}^{1,2,3}$\thanks{JDM is supported by the
    Swiss National Science Foundation (SNSF) under grant
    200021-130359.  YW is supported in part by the Center for
    Biomedical Imaging (CIBM) of the Geneva and Lausanne Universities, 
    EPFL, and the Leenaards and Louis-Jeantet foundations, and in part by
    the SNSF under grant PP00P2-123438.}}
\address{${}^1$Institute of Electrical Engineering, Ecole Polytechnique
  F{\'e}d{\'e}rale de Lausanne (EPFL), Switzerland\\
  ${}^2$Institute of Bioengineering, Ecole Polytechnique
  F{\'e}d{\'e}rale de Lausanne (EPFL), Switzerland\\
  ${}^3$Department of Radiology \& Medical Informatics, University of
  Geneva (UniGE), Switzerland
}
\begin{document}
%
\maketitle
%
\begin{abstract}
Radio interferometry is a powerful technique for astronomical imaging.  The theory of \csfirst\ has been applied recently to the ill-posed inverse problem of recovering images from the measurements taken by radio interferometric telescopes.  We review novel \cs\ radio interferometric imaging techniques, both at the level of acquisition and reconstruction, and discuss their superior performance relative to traditional approaches.  In order to remain as close to the theory of \cs\ as possible, these techniques necessarily consider idealised interferometric configurations.  To realise the enhancement in quality provided by these novel techniques on real radio interferometric observations, their extension to realistic interferometric configurations is now of considerable importance.  We also chart the future direction of research required to achieve this goal.
\end{abstract}
%
\section{Introduction}

Aperture synthesis interferometry is a powerful technique in radio astronomy, providing astronomical images at angular resolution and sensitivity that are otherwise inaccessible (see \eg\ \cite{thompson:2001}).  Arrays of telescopes track a source as it traverses the sky to synthesise the aperture of a telescope of much larger size than the individual telescopes comprising the array.  Radio interferometric arrays directly probe the Fourier plane associated with the image tangent to the celestial sphere at the source position.  As the source traverses the sky, interferometric measurements, termed visibilities, are generated on an elliptical track of positions in the Fourier plane.  Visibility coverage in the Fourier plane is therefore incomplete.  Consequently, recovering interferometric images from measured visibilities requires solving an ill-posed deconvolution problem.

The recovery of interferometric images has traditionally been tackled by the so-called CLEAN algorithm \cite{hogbom:1974}, which is a local matching pursuit approach.  Recently, however, the theory of \csfirst\ \cite{candes:2006a,donoho:2006} has been applied to interferometric imaging with considerable success, both at level of acquisition and reconstruction \cite{wiaux:2009:cs,wiaux:2009:ss,mcewen:riwfov}.  Furthermore, \cs\ techniques have been developed to successfully extract astronomical signals of interest from interferometric observations corrupted by background contributions \cite{wiaux:2010:csstring}.  First steps in these directions necessarily involved the study of idealised interferometric configurations to remain as close to the theory of \cs\ as possible.  With a new generation of interferometric telescopes under design and construction, radio interferometry will continue to play a fundamental role in astronomy.  The extension of \cs\ techniques to more realistic interferometric configurations is therefore now of paramount importance, in order to realise the enhancement in quality that these techniques provide on real observations made by radio interferometric telescopes.

In this article we review the current state-of-the-art of \cs\ imaging techniques for radio interferometry.  We then chart the future direction that this research must take to be applicable to real radio interferometric telescopes.  In \sectn{\ref{sec:background}} we review radio interferometry concisely and pose the interferometric imaging inverse problem.  We discuss the \cs\ techniques applied to this problem in \sectn{\ref{sec:imaging}}, before charting the future direction of research in \sectn{\ref{sec:future}}.

\section{Interferometric inverse problem}
\label{sec:background}

\begin{figure}[tb]
\centering
\includegraphics[trim = 0cm 4cm 0cm 0cm, clip, width=50mm]{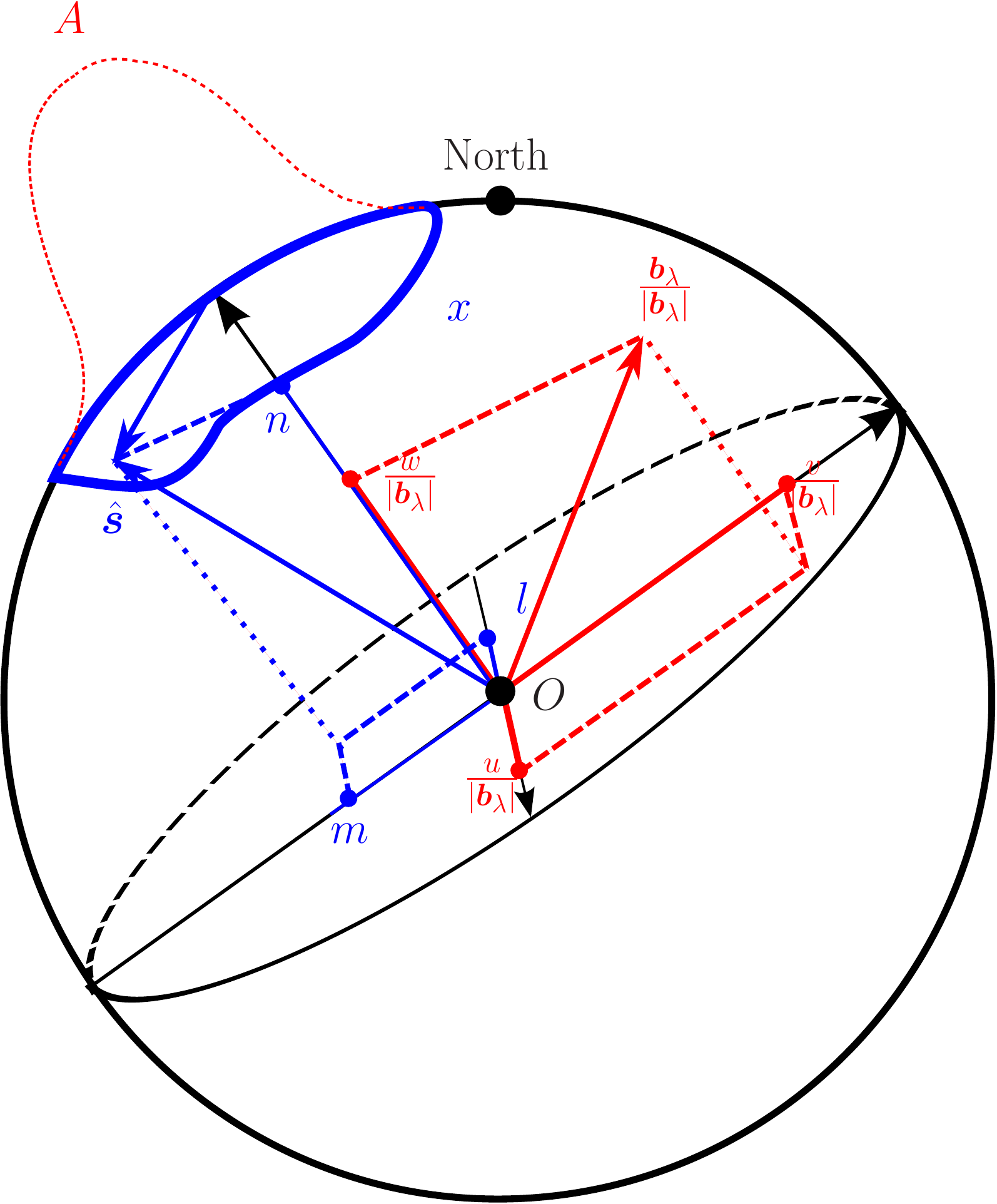}
\caption{Coordinate system for interferometric imaging (credit: \cite{wiaux:2009:ss}).}
\label{fig:notation}
\end{figure}

The field-of-view (\fov) observed by an interferometer is limited by the primary beam of the telescope $\beam$.  We denote the sky intensity to be imaged by \im\ and consider a coordinate system centred on the pointing direction of the telescope $\sao$, where a point \sa\ on the celestial sphere is defined by its coordinates $(\lx,\mx,\nx)$, noting $\nx \equiv \nx(\lxvect) = ( 1-\| \lxvect \|_2 ^2)^{1/2}$, for $\lxvect=(\lx,\mx)$ and $\| \lxvect \|_2 ^2 = \lx^2 + \mx^2$.  This configuration is illustrated in \fig{\ref{fig:notation}}.  The complex visibility measured by the correlation of the incoming electric fields of the observed source for each telescope pair is given by \cite{thompson:2001}
\begin{equation}
\label{eqn:vis3}
\vis(\buvect, \bw) = \int_\disk 
\beam(\lxvect) \: \im(\lxvect) \: \chirpfull \:
\exp{-\img 2 \pi \buvect \cdot \lxvect } \:
\frac{\dx^2 \lxvect}{\nx (\lxvect)}
\spcend ,
\end{equation}
where the measured visibility depends only on the relative positions between telescope pairs, denoted in units of wavelengths by the baseline vector $\buvectfull_{\lambda}=(\bu,\bv,\bw)$, with \mbox{$\buvect=(\bu,\bv)$}, and the so-called \wterm\ is given by
\begin{equation*}
\label{eqn:chirpfull}
\chirpfull \equiv \exp{
\img 2 \pi \bw \bigl ( 1 - \sqrt{1-\| \lxvect \|_2 ^2} \bigr )
}
\spcend .
\end{equation*}
Visibilities are acquired for non-zero $\bw$ if the baseline of the telescope pair has a component in the pointing direction of the telescope.

Typically small \fov\ assumptions are made, in which case the approximations $\nx (\lxvect) \simeq 1$ and $\chirpfull \simeq 1$ reduce \eqn{\ref{eqn:vis3}} to a standard Fourier transform.  However, if these assumptions are relaxed, $\nx^{-1}(\lxvect)$ may instead be absorbed into the beam but a \wterm\ remains.  The \wterm\ gives rise to the spread spectrum phenomenon, which is a serendipitous property of the acquisition, enhancing the performance of \cs\ techniques considerably.

To recover the source image from incomplete visibility measurements, we pose the inverse problem \eqn{\ref{eqn:vis3}} in a discrete setting.  The following formulation is restricted to constant \bw, thus restricting visibilities to a Fourier plane.\footnote{The restriction to constant \bw\ allows one to discard considerations related to specific interferometer configurations and to study the spread spectrum phenomenon at light computational load \cite{wiaux:2009:ss}.}
We measure \nmeas\ discrete visibilities denoted by the vector \visvect, falling on a grid of frequencies of size $\ndim = \ndim^{1/2} \times \ndim^{1/2}$, which are related to the \ndim\ source image pixels, denoted by the vector \imvect, through the linear system
\begin{equation}
\label{eqn:vis_linear_plane}
\visvect = \sensmatp \impvect + \nimvect
\spcend ,
\end{equation}
with
\begin{equation*}
\sensmatp  = \opmask \, \opfourier \, \opchirp \, \opbeam
\spcend ,
\end{equation*}
where the vector \nimvect\ represents noise corrupting the measurements.  The $\nmeas \times \ndim$ measurement matrix \sensmatp\ identifies the complete linear relation between the signal and the visibilities.
The matrix $\opbeam$ is the diagonal matrix implementing the primary beam.  The matrix $\opchirp$ is the diagonal matrix implementing the modulation by the \wterm, which reduces to the identity if small \fov\ assumptions are made.  The unitary matrix $\opfourier$ implements the discrete Fourier transform.  The $\nmeas \times \ndim$ matrix $\opmask$ is the rectangular binary matrix implementing the mask that characterises visibility coverage.
Given incomplete visibility coverage with $\nmeas<\ndim$, \eqn{\ref{eqn:vis_linear_plane}} defines an ill-posed inverse problem.

\section{Compressed sensing\\for interferometric imaging}
\label{sec:imaging}

\Cs\ reconstruction techniques have been applied to interferometric imaging recently \cite{wiaux:2009:cs}, outperforming the standard CLEAN algorithm.  The spread spectrum phenomenon which arises naturally in radio interferometry, provides a further enhancement in reconstruction quality \cite{wiaux:2009:ss}.  \mbox{Extensions} to wide field imaging provide another enhancement \cite{mcewen:riwfov}.  We review these works in this section.

\subsection{Sparsity for interferometric image reconstruction}

The theory of \cs\ acknowledges that a large variety of signals in Nature are sparse.  In this context, the interferometric ill-posed inverse problem can be regularised by a sparsity or compressibility prior.  This results in an optimisation problem called Basis Pursuit (BP), which consists of minimising the $\el_1$-norm of the coefficients $\vect{\alpha}$ of the signal represented in a sparsity basis $\sparmat$ (the $\ndim \times \ndim$ matrix with columns containing the $\sparatom$ vectors of the sparsity basis), under a constraint on the $\el_2$-norm of the residual noise:
\begin{equation*}
\label{eqn:min_bp}
\min_{\vect{\alpha}} \| \vect{\alpha} \|_{1} \:\: \mbox{such that} \:\:
\| \vect{y} - \sensmat \sparmat \vect{\alpha}\|_2 \leq \epsilon
\spcend .
\end{equation*}
Recall that the $\el_1$-norm is given by $\|\vect{\alpha} \|_1=\sum_{i=1}^{\ndim} | \alpha_i |$, while the $\el_2$-norm is given by $\|\vect{\alpha} \|_2=(\sum_{i=1}^{\ndim}  | \alpha_i |^2)^{1/2}$.  The constraint $\epsilon$ may be related to a residual noise level estimator \cite{wiaux:2009:cs}.  The BP problem may be solved by the application of non-linear, iterative convex optimisation algorithms.  If the solution of the optimisation problem is denoted $\vect{\alpha}^\star$, then the signal is reconstructed through the synthesis $\vect{x}^{\star} = \sparmat \vect{\alpha}^\star$.

\begin{figure}[tb]
\centering
\subfigure[Typical simulation]{\includegraphics[width=40mm]{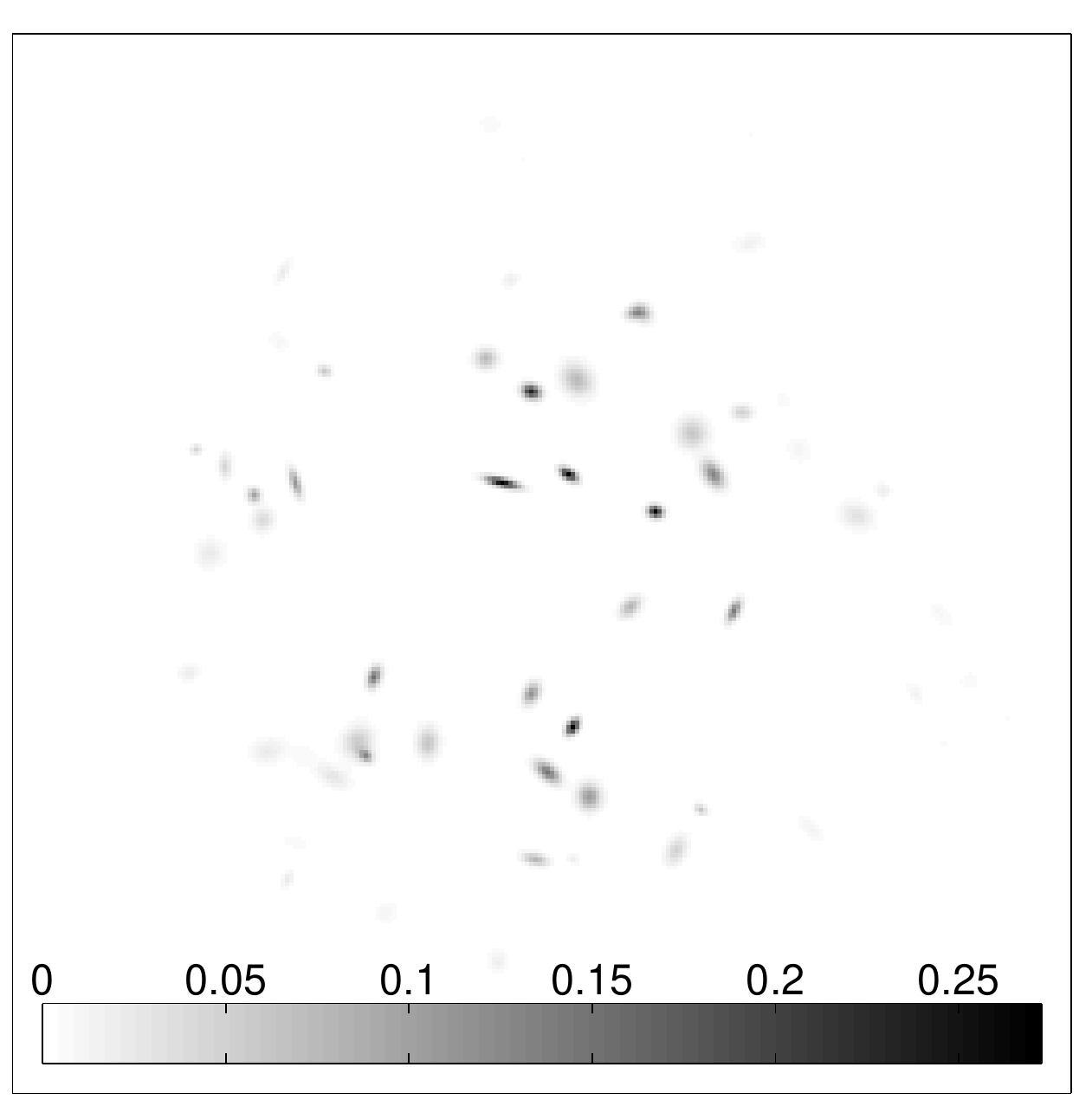}} 
\subfigure[Reconstruction performance]{\includegraphics[width=40mm]{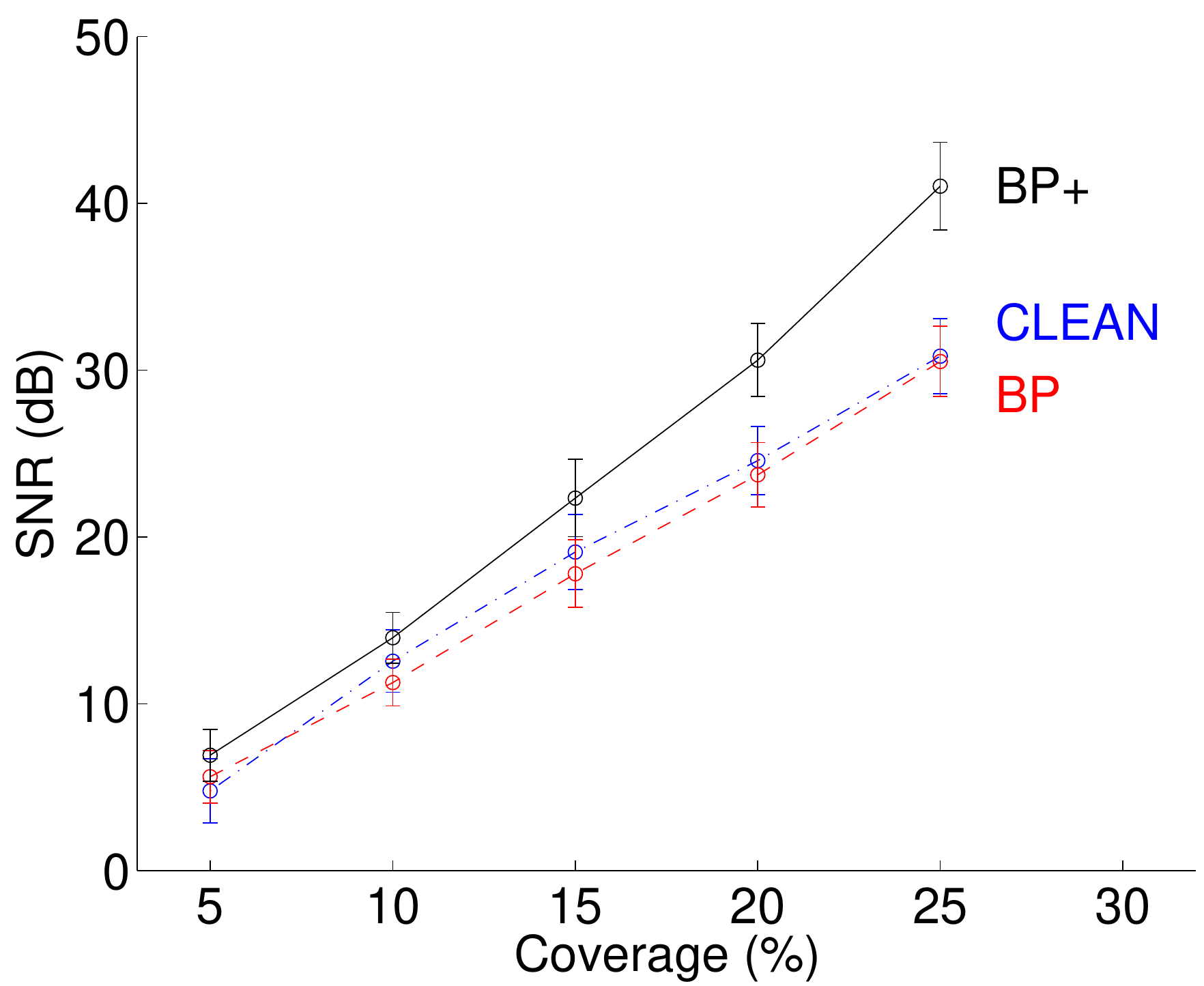}} 
\caption{BP, BP+ and CLEAN reconstruction performance on elongated Gaussian profile signals (credit: \cite{wiaux:2009:cs}).}
\label{fig:cs}
\end{figure}

Under typical small \fov\ assumptions, solving the BP problem provides similar reconstruction quality to multi-scale versions of the CLEAN algorithm.  However, the versatility of this framework allows the easy addition of many useful priors, such as the positivity prior leading to BP+ reconstructions of superior quality to CLEAN \cite{wiaux:2009:cs}.  A typical simulation used to verify this result and the reconstruction quality of various imaging techniques, as a function of the percentage of visibilities measured, are shown in \fig{\ref{fig:cs}}.  The Dirac sparsity basis was used in this work.

\subsection{Spread spectrum phenomenon}
\label{sec:imaging:ss}

If small \fov\ assumptions are not made, then the \wterm\ modulation \opchirp\ no longer reduces to the identity.  
Beyond sparsity, \cs\ demonstrates that the incoherence between the measurement and sparsity bases is a key criterion driving the performance of \cs\ based reconstructions.  In the interferometric setting the measurement basis is essentially the Fourier basis, hence coherence is given by the maximum modulus of the Fourier coefficient of the sparsity basis vectors.  Consequently, an operation that acts to reduce the maximum Fourier coefficient, reduces the coherence and thus improves the quality of \cs\ reconstructions.  The \wterm\ corresponds to a norm-preserving convolution in the Fourier plane, spreading the spectrum of the sparsity basis vectors, and achieves exactly this.  The Dirac basis considered previously is already maximally incoherent, hence this spread spectrum phenomenon is not expected to improve performance for this case.  However, for alternative bases we expect an enhancement.

\begin{figure}[tb]
\centering
\subfigure[Typical simulation]{\includegraphics[width=40mm]{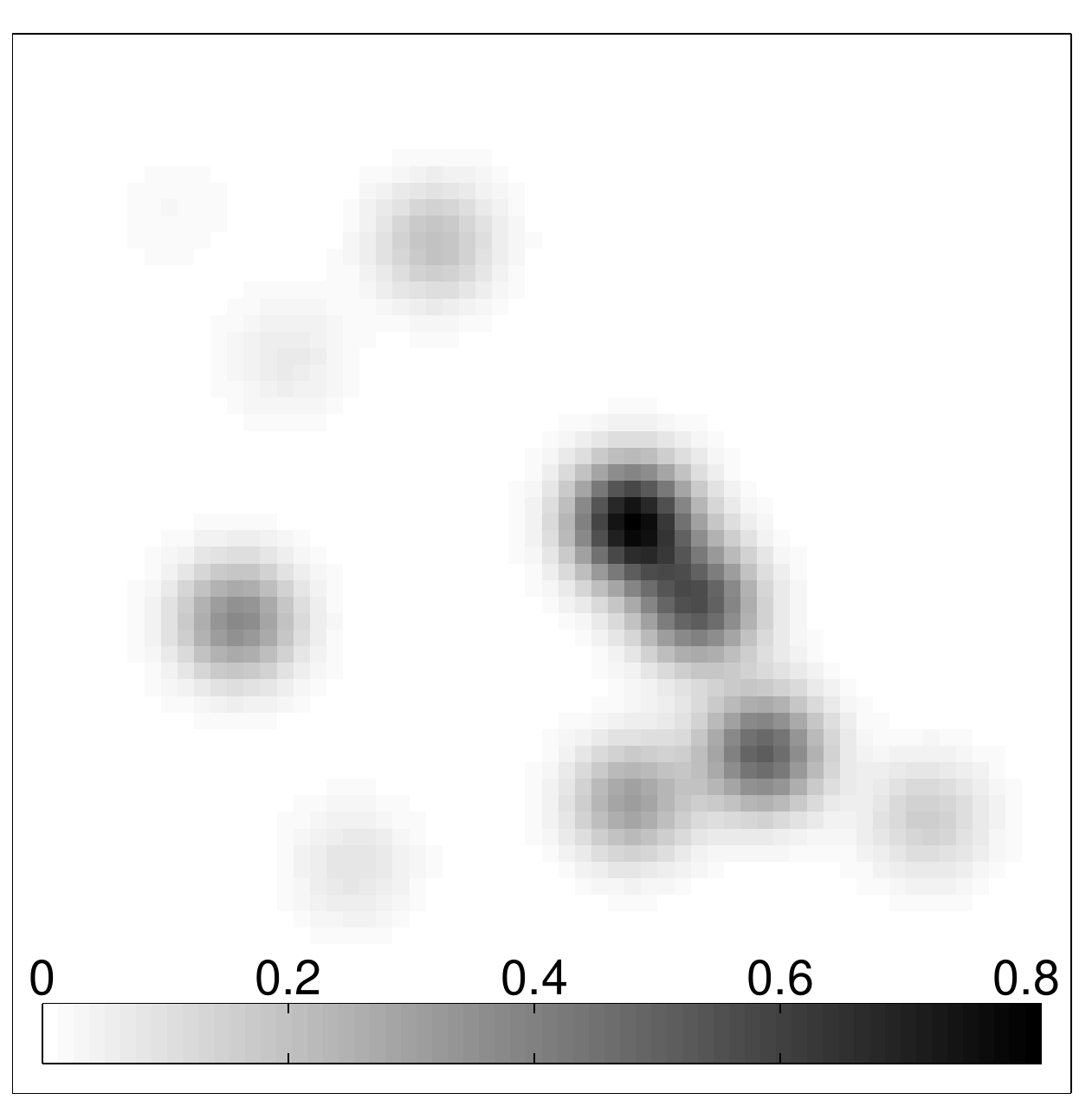}} 
\subfigure[Reconstruction performance]{\includegraphics[width=40mm]{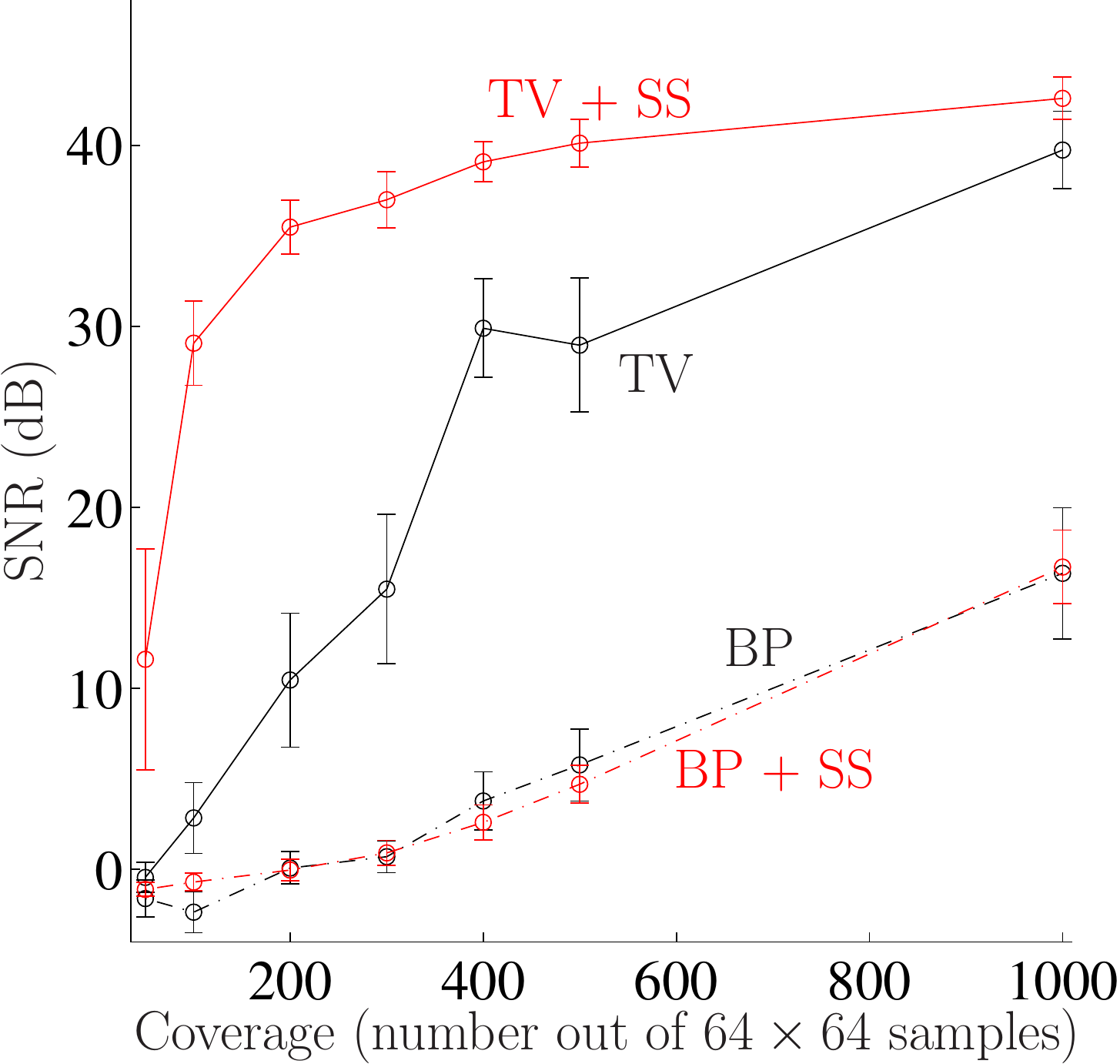}} 
\caption{BP and TV reconstruction performance on Gaussian profile signals, in the absence and presence of the spread spectrum (SS) phenomenon (credit: \cite{wiaux:2009:ss}).}
\label{fig:ss}
\end{figure}

Many natural signals are also sparse or compressible in the magnitude of their gradient, in which case the Total Variation (TV) minimisation problem applies \cite{candes:2006a}.  The TV problem involves replacing the \mbox{$\el_1$-norm} sparsity prior in the BP problem with a prior on the TV norm of the signal itself:
\begin{equation*}
\label{eqn:min_tv}
\min_{\vect{x}} \| \vect{x} \|_{\rm TV} \:\: \mbox{such that} \:\:
\| \vect{y} - \sensmat \vect{x}\|_2 \leq \epsilon
\spcend ,
\end{equation*}
where the TV norm is defined by the $\el_1$-norm of the gradient of the signal $\| \vect{x} \|_{\rm TV} = \| \nabla x \|_1$.  The TV problem may also be solved by the application of non-linear, iterative convex optimisation algorithms, illustrating another example of the versatility of the framework.  The signal is directly recovered from the solution to the optimisation problem.  Recovering interferometric images by solving the TV problem has been shown to achieve very good reconstruction quality and, moreover, is enhanced considerably by the spread spectrum phenomenon \cite{wiaux:2009:ss}.  A typical simulation used to study this effect and reconstruction quality for BP (with Dirac sparsity) and TV approaches, in the absence and presence of the spread spectrum phenomenon, are shown in \fig{\ref{fig:ss}}.

\subsection{Wide field-of-view}

Incorporating wide \fov\ considerations in interferometric imaging is of increasing importance for the next generation of radio interferometric telescopes.  \Cs\ techniques for interferometric imaging have been extended to a wide \fov\ recently \cite{mcewen:riwfov}, recovering images in the spherical coordinate space in which they naturally live.  The effectiveness of the spread spectrum phenomenon is enhanced when going to a wide \fov\ since the \wterm\ evolves to include higher frequency content, while sparsity is promoted by recovering images directly on the sphere.  Both of these properties act to improve the quality of reconstructed interferometric images. 

The wide \fov\ ill-posed inverse problem is defined by 
\begin{equation}
\label{eqn:vis_linear_sphere}
\visvect = \sensmats \imsvect + \nimvect
\spcend ,
\end{equation}
with 
\begin{equation*}
\sensmats  = \opmask \, \opfourier \, \opchirp \, \opbeam
\, \opproj
\spcend ,
\end{equation*}
where the vector \imsvect\ is the interferometric image defined on the celestial sphere (rather than a tangent plane).  The measurement operator on the sphere $\sensmats$ simply consists of augmenting the operator on the plane $\sensmatp$ by prepending a projection $\opproj$ from the sphere to the plane, which incorporates a convolutional regridding.  Careful consideration is given to samplings on the sphere and plane to ensure that the planar grid is sampled sufficiently to accurately represent the projection of a  band-limited signal defined on the sphere.  The BP and TV problems are then solved to recover interferometric images on the sphere.  This approach has been shown to perform very well, with the quality of recovered spherical interferometric images superior to planar reconstructions \cite{mcewen:riwfov}.  A typical simulation used to study this effect and reconstruction quality for BP (with Dirac sparsity) and TV approaches both on the planar and spherical settings, in the absence and presence of the spread spectrum phenomenon, are shown in \fig{\ref{fig:riwfov}}.

\newlength{\reconplotwidth}
\setlength{\reconplotwidth}{40mm}

\begin{figure*}[tb]
\centering
\subfigure[Typical simulation]{\includegraphics[clip=,viewport=180 30 600 480,width=\reconplotwidth]{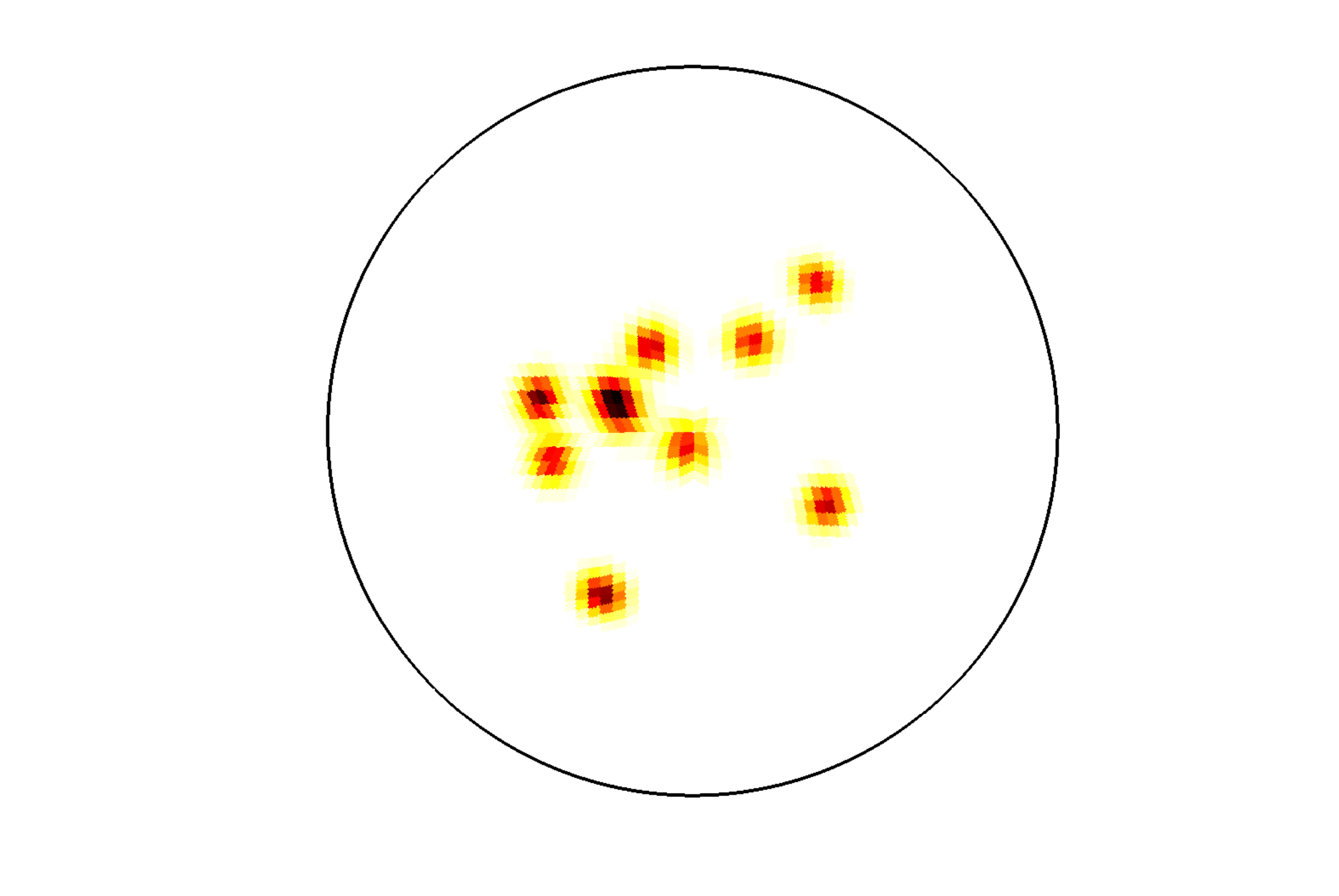}}\quad\quad
\subfigure[BP reconstruction performance]{\includegraphics[clip=,viewport=130 0 610 480,width=\reconplotwidth]{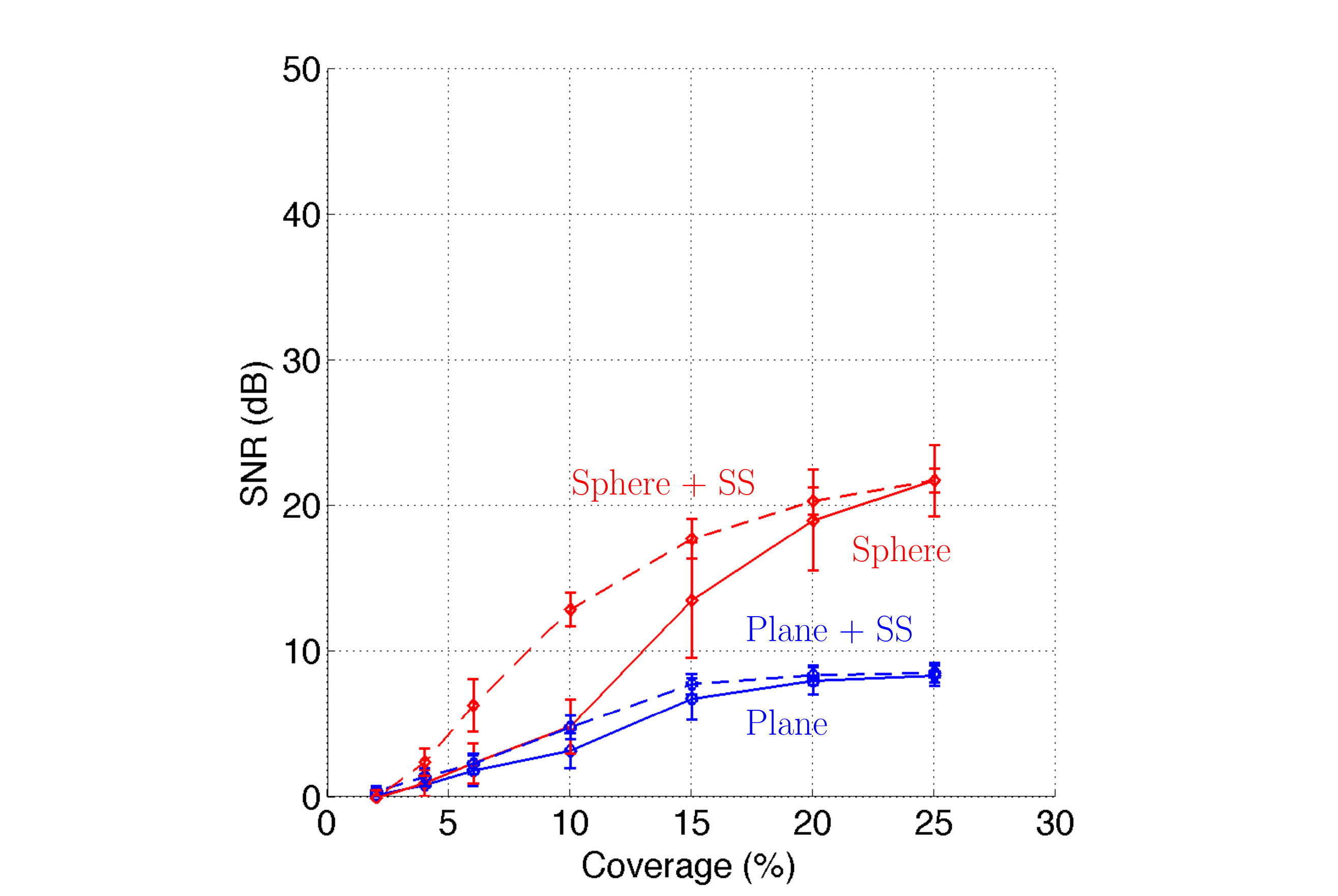}}\quad\quad
\subfigure[TV reconstruction performance]{\includegraphics[clip=,viewport=130 0 610 480,width=\reconplotwidth]{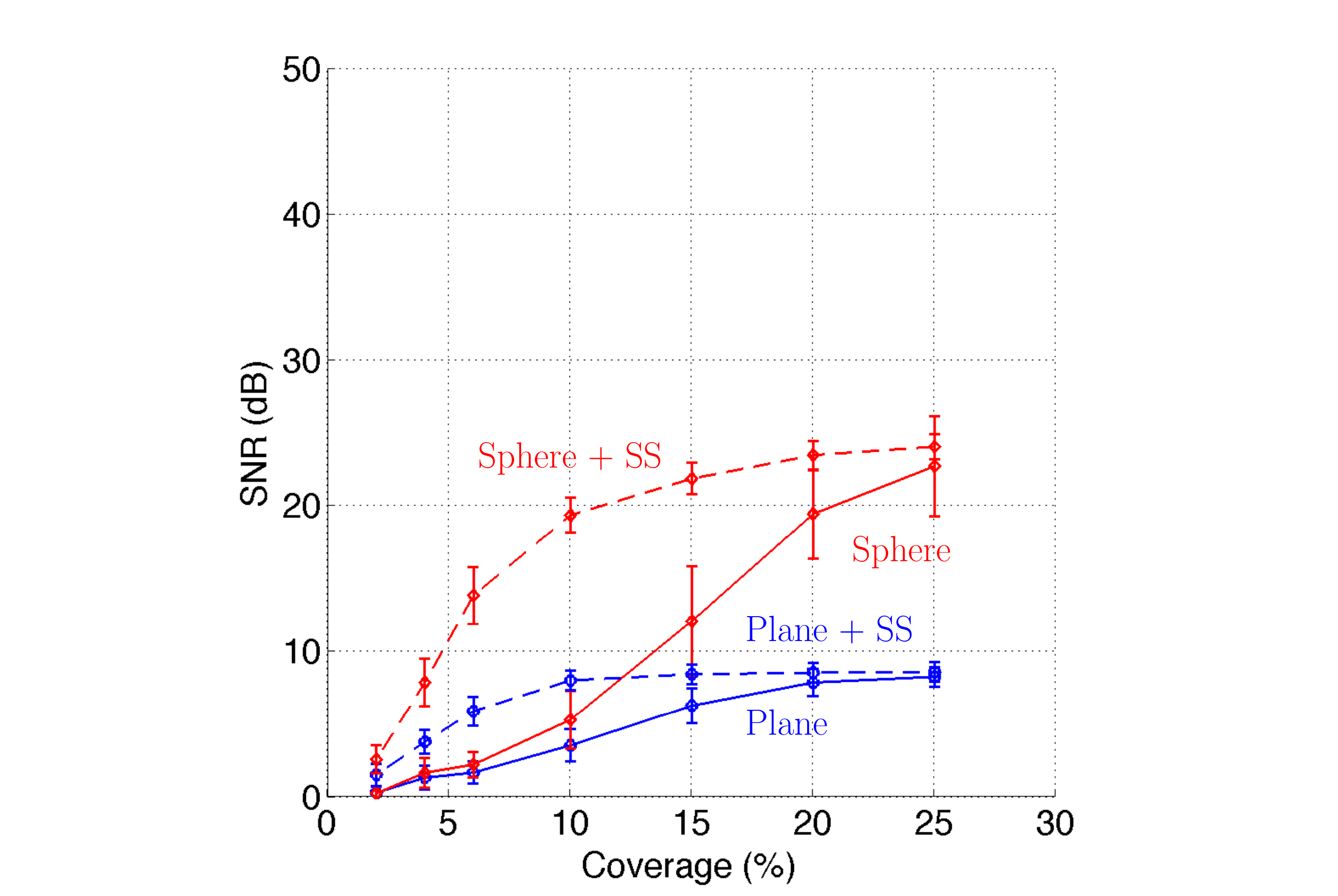}}
\caption{BP and TV reconstruction performance on Gaussian profile signals on the plane and sphere, in the absence and presence of the spread spectrum (SS) phenomenon (credit: \citet{mcewen:riwfov}).}
\label{fig:riwfov}
\end{figure*}

\section{Future direction}
\label{sec:future}

The \cs\ techniques developed for interferometric imaging \cite{wiaux:2009:cs,wiaux:2009:ss,mcewen:riwfov} remain somewhat idealised.  Random visibility coverage in $(\bu,\bv)$ is assumed, with the spatial frequencies probed by the interferometer also assumed to fall on discrete grid points.  Furthermore, to study the spread spectrum phenomenon a constant $\bw$ is assumed.  These restrictions have been necessary to remain as close to the theory of \cs\ as possible during the development and evaluation of interferometric imaging techniques.  However, now that the effectiveness of these techniques has been demonstrated, it is important to adapt them to realistic interferometric configurations.

In practice a varying \bw\ will not alter the essence of the spread spectrum phenomenon due to its realisation through the convolution operator; however, this effect must be studied.  The $\bw$-projection algorithm \cite{cornwell:2008:w} may be applied to reduce the computational cost in this context.

Realistic visibility coverages in all components $(\bu,\bv,\bw)$ must also be considered by simulating observations made by real interferometer configurations.  Furthermore, to integrate \cs\ imaging techniques into the data processing pipelines of radio interferometric telescopes, it is necessary to consider the continuous visibilities measured by an interferometer before they are gridded (so that imaging can be integrated into the iterative self-calibration of telescopes).

In general, \cs\ addresses imaging by optimising both reconstruction and acquisition, while current applications in interferometry have essentially focused on reconstruction only.  The possibility of optimising the configuration of interferometers to enhance the spread spectrum phenomenon \cite{wiaux:2009:ss} for \cs\ reconstruction is an exciting avenue of research at the level of acquisition, contrary to the traditional approach that attempts to minimise this effect.  In addition, direction dependent beam effects may also provide an alternative source of the spread spectrum phenomenon.  

In conclusion, with a new generation of radio interferometric telescopes under design and construction, such as the Square Kilometer Array (SKA), the development of novel interferometric techniques is of paramount importance.  It is essential that the \cs\ interferometric imaging techniques developed recently are adapted to realistic settings to ensure that the fidelity of reconstructed images keeps pace with the capabilities of new instruments.

\setlength{\bibsep}{0.5mm}
\renewcommand{\bibsection}{\section{\normalsize References}}
\small
\bibliographystyle{IEEEbib}
\bibliography{bib}

\end{document}